\documentclass[prd,twocolumn,showpacs,superscriptaddress,nofootinbib,floatfix,showkeys,10pt]{revtex4-2}
\usepackage{bm}
\usepackage{times}
\usepackage{braket}
\usepackage{amsfonts,amssymb,stmaryrd,latexsym,amsmath}
\usepackage[usenames,dvipsnames]{color}
\usepackage{epsfig}
\usepackage{slashed}
\usepackage{hyperref}
\usepackage{subfigure}
\usepackage{graphics}
\usepackage{slashed}
\usepackage{orcidlink}
\usepackage{multirow}
\usepackage{dashrule}
\allowdisplaybreaks[1]

\definecolor{nicered}{rgb}{0.7,0.1,0.1}
\definecolor{nicegreen}{rgb}{0.1,0.5,0.1}
\definecolor{emph}{rgb}{1,0,0}
\definecolor{doub}{rgb}{0.7,0.2,1.0}
\definecolor{navyblue}{RGB}{0, 110, 184}

\hypersetup{colorlinks,citecolor=nicegreen,linkcolor=nicered,urlcolor=navyblue}

\newcommand{\clabel}[2][]{#2}

\begin{document}

	
\title{Virtual states in the coupled-channel problems with an improved complex scaling method} 

 \author{Yan-Ke Chen\,\orcidlink{0000-0002-9984-163X}}\email{chenyanke@stu.pku.edu.cn}
\affiliation{School of Physics, Peking University, Beijing 100871, China}

\author{Lu Meng\,\orcidlink{0000-0001-9791-7138}}\email{lu.meng@rub.de}
\affiliation{Institut f\"ur Theoretische Physik II, Ruhr-Universit\"at Bochum,  D-44780 Bochum, Germany }

\author{Zi-Yang Lin\,\orcidlink{0000-0001-7887-9391}}\email{lzy\_15@pku.edu.cn}
\affiliation{School of Physics, Peking University, Beijing 100871, China}

\author{Shi-Lin Zhu\,\orcidlink{0000-0002-4055-6906}}\email{zhusl@pku.edu.cn}
\affiliation{School of Physics and Center of High Energy Physics,
Peking University, Beijing 100871, China}

\begin{abstract}

We improve the complex scaling method (CSM) to obtain virtual states, which is challenging in the conventional CSM.  Our approach solves the Schr\"odinger equation in momentum space as an eigenvalue problem by choosing flexible contours. It proves to be highly effective in identifying the poles across different Riemann sheets in multichannel scatterings. This approach is more straightforward and efficient than employing the Lippmann-Schwinger equation and using a root-finding algorithm to search for the zeros of the Fredholm determinant. This advancement significantly extends the capabilities of the CSM in accurately characterizing the virtual states in quantum systems.
\end{abstract}
 
\maketitle


\section{Introduction}

\begin{figure}[htbp]
	\centering
	\includegraphics[width=0.475\textwidth]{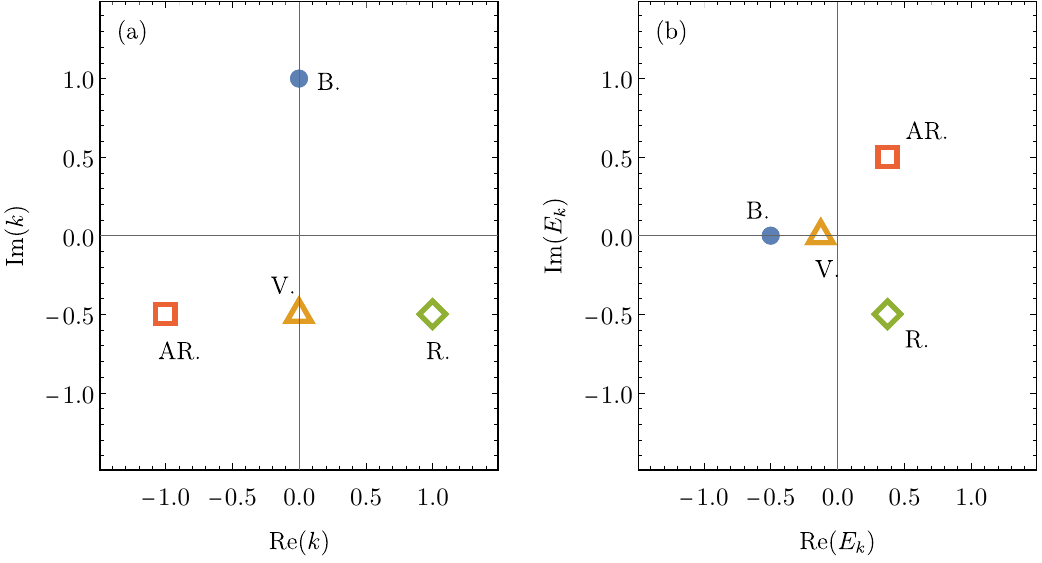}
	\caption{The $T$ matrix poles in Eq.~\eqref{eq:4poles} corresponding to the bound state (B), virtual state (V), resonance (R), and anti-resonance (AR) in (a) the momentum plane and (b) the energy plane. The solid and hollow markers indicate the poles on the RS-I and -II, respectively.}
	\label{fig:pole_position_illustration}
\end{figure}

\begin{figure}[htbp]
	\centering
	\includegraphics[width=.35\textwidth]{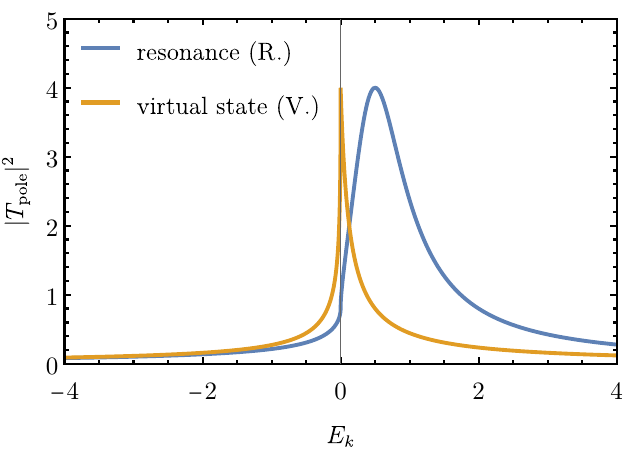}
	\caption{The contributions of the virtual state and resonance in Eq.~\eqref{eq:4poles} to the physical $T$ matrix estimated by Eq.~\eqref{eq:Tapprox}, with couplings set to $g_R=g_V=1$. The resonance enhances the $T$ matrix at $E=\mathrm{Re}(E_{k,r})$, while the virtual state causes an enhancement exactly at the threshold.}
	\label{fig:pole_enhancemant}
\end{figure}

In the context of a two-body scattering, bound states, virtual states, resonances, and antiresonances are poles of the $S$ matrix or $T$ matrix. The scattering momentum $k=\sqrt{2\mu E_k}$ is proportional to the square root of the scattering energy $E_k$, \clabel[symbolexplain]{where $\mu$ is the reduced mass}. The $S$ matrix or $T$ matrix becomes a double-valued function in the complex $E_k$ plane. For a single-channel scattering, the Riemann sheets (RSs) with positive and negative imaginary parts of $k$ are called RS-I (``physical" sheet) and RS-II (``unphysical sheet"), respectively~\cite{Badalian:1981xj}. The poles of the $T$ matrix are classified as follows\footnote{In some literature, the $\gamma_r>\kappa_r$ resonance states are also defined as virtual states.} ($\kappa_r, \gamma_{r, b, v}>0$)~\cite{aoyama2006complex}:
$$
\begin{array}{lll}\text { bound states: } & k_B=i \gamma_b & \text{(RS-I)}, \\
\text { virtual states: } & k_{V}=-i \gamma_{v} & \text{(RS-II)}, \\ 
\text { resonances: } & k_R=\kappa_r-i \gamma_r & \text{(RS-II)}, \\ 
\text { antiresonances: } & k_{A R}=-\kappa_r-i \gamma_r & \text{(RS-II)}.\end{array}
$$
\clabel[poleillustration]{For illustration, we assume that the $T$ matrix for a two-body scattering process has four following poles:
\begin{equation}\label{eq:4poles}
    k_B=-i,~k_V=-0.5i,~k_R=1-0.5i,~k_{AR}=-1-0.5i,
\end{equation}
and the reduced mass is taken to be dimensionless $\mu=1$. The distribution of these poles in the momentum plane and energy plane is shown in Fig.~\ref{fig:pole_position_illustration}. In the vicinity of the pole, the $T$ matrix is dominated by
\begin{equation}\label{eq:Tapprox}
T_{\mathrm{pole}}(E) \sim \frac{g_{\mathrm{pole}}}{E-E_{\mathrm{pole}}}.
\end{equation}
Therefore, the virtual states and resonances will enhance the physical $T$ matrix if they are very close to the physical sheet. To illustrate this, we use the approximation in Eq.~\eqref{eq:Tapprox} to estimate the contributions of the virtual state and resonance in Eq.~\eqref{eq:4poles} to the physical $T$ matrix, as depicted in Fig.~\ref{fig:pole_enhancemant}. It is evident that the resonance enhancement appears above the threshold at $E=\mathrm{Re}(E_{k,r})$. The virtual state pole leads to an enhancement exactly at the threshold and amplifies the ``threshold cusp".} One can consult Refs.~\cite{Dong:2020hxe,Meng:2022ozq} for more detailed discussions on the relations between near-threshold poles and line shapes.

For the past decades, numerous enhancements observed in nuclear physics and particle physics experiments could be attributed to poles near the physical Riemann sheet. A great number of exotic hadron candidates have been discovered in the past decades, such as $X(3872)$~\cite{Belle:2003nnu}, $T_{cc}^+$~\cite{LHCb:2021vvq, LHCb:2021auc}, and hidden-charm pentaquarks~\cite{LHCb:2015yax,LHCb:2019kea,LHCb:2022jad}. These states lie in close proximity to hadron-hadron thresholds, presenting challenges for both experimental analyses and theoretical interpretations. The nature of many exotic hadronic states remains elusive. Some of them can be interpreted as bound states, virtual states, or resonances of two hadrons considering possible coupled-channel effects. One can refer to Refs.~\cite{Chen:2016qju, Guo:2017jvc,Guo:2019twa,Brambilla:2019esw,Liu:2019zoy,Meng:2022ozq, Chen:2022asf} for recent reviews and other theoretical interpretations. In nuclear physics, it is also very crucial to identify the poles to understand some atomic nuclei, such as the halo nuclei~\cite{Hammer:2019poc,Hammer:2017tjm}. Therefore, obtaining information on the poles of the $S$ matrix is of great significance.

In principle, the poles of a two-body scattering $S$ matrix or $T$ matrix can be determined by finding the zeros of the Fredholm determinant of the Lippmann-Schwinger equation (LSE). One can adopt some root-finding algorithms or optimization algorithms in the entire complex $E$ plane to find the zeros of the Fredholm determinant. However, such a process is usually iterative and, thus, time consuming. Alternatively, for bound states, a more efficient approach is to expand the Hamiltonian with some square-integrable bases and transform the task into a more tractable eigenvalue problem. 

The complex scaling method (CSM) proves to be a highly effective technique to directly obtain the resonances~\cite{Aguilar:1971ve, Balslev:1971vb, aoyama2006complex}. The resonances can be solved like bound states via the square-integrable basis expansion by introducing a simple complex scaling rotation. The CSM has been widely used in atomic and molecular physics, nuclear physics, and hadronic physics to investigate the resonances~\cite{Myo:2001wn,aoyama2006complex,Dote:2012bu,Carbonell:2013ywa,Papadimitriou:2014bfa,Maeda:2018xcl,Myo:2020rni,Yang:2022bfu,Wang:2022yes}. Recently, the CSM in momentum space has been adopted to investigate the exotic hadrons~\cite{Cheng:2022qcm, Lin:2022wmj, Lin:2023dbp}.

\begin{figure}[htbp]
  \includegraphics[width=.4\textwidth]{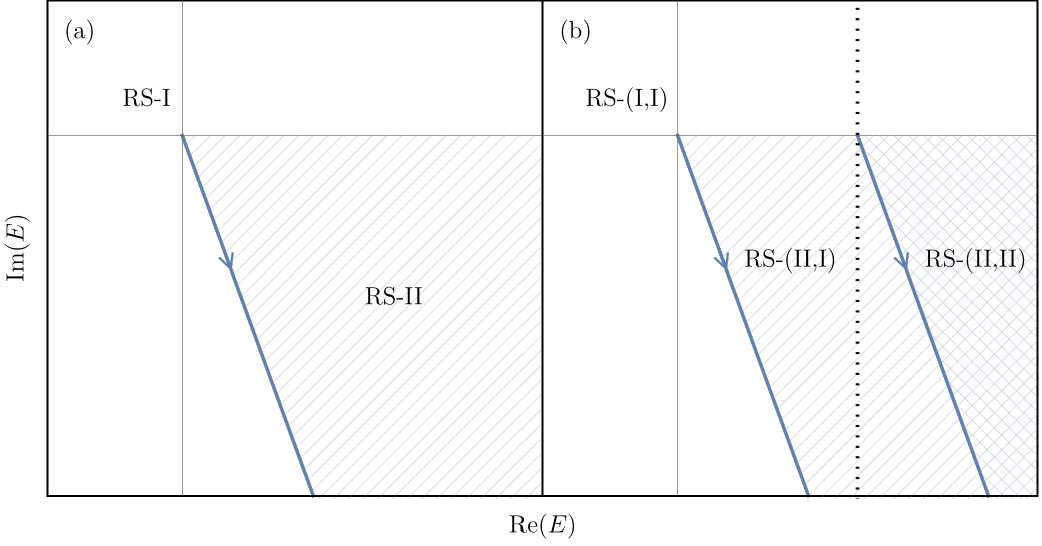}
  \caption{The regions of the Riemann sheets that the CSM can explore in (a) a single-channel and (b) a double-channel problem.}
  \label{fig:csm_intro}
\end{figure}

However, in practical applications, the regions of the Riemann sheets that the CSM can probe is severely limited. For a single-channel problem, the CSM can cover only a portion of the fourth quadrant of the RS-II, as illustrated by the shaded region in Fig.~\ref{fig:csm_intro} (a).  It remains challenging for the CSM to directly obtain the virtual states. The virtual states can be obtained through several other methods, such as the analytical continuation in the coupling constant \cite{tanaka1999unbound}, the Jost function method \cite{Masui:2000mug}, and extracting from the level density calculated in the CSM \cite{Odsuren:2017urx}. But these methods are very complicated and need to be more intuitive. In the double-channel scattering, there are more Riemann sheets. Only a small portion is reachable via CSM, as illustrated in Fig.~\ref{fig:csm_intro}(b).

In this work, we propose an improved complex scaling method (ICSM) to obtain the virtual states. The ICSM is similar to the conventional CSM in obtaining the bound state and resonance poles. This approach also exhibits advantages when dealing with the multichannel scattering problems, which enables the identification of the poles previously inaccessible to the conventional CSM. We apply this method to several different potentials as examples, including the separable potential, the HAL QCD $DD^*$ potential \cite{Lyu:2023xro}, and the high-quality $NN$ Reid93 potential \cite{Stoks:1994wp}, and compare the results with those from the LSE and CSM.

This paper is organized as follows. In Sec.~\ref{formalism}, we introduce the ICSM framework. In Secs.~\ref{sec:single-channel} and \ref{sec:multichannel}, we extract the poles in the single-channel and multichannel scatterings, respectively. We use various potential models as examples and compare the results of the ICSM and other methods. We provide a summary and discussion in Sec.~\ref{sec:summary} .

\section{formalism}\label{formalism}

\begin{figure*}[tbp]
	\centering

	\subfigure
	{
	\begin{minipage}[c]{0.95\linewidth}
		\centering
		\includegraphics[width=1\linewidth]{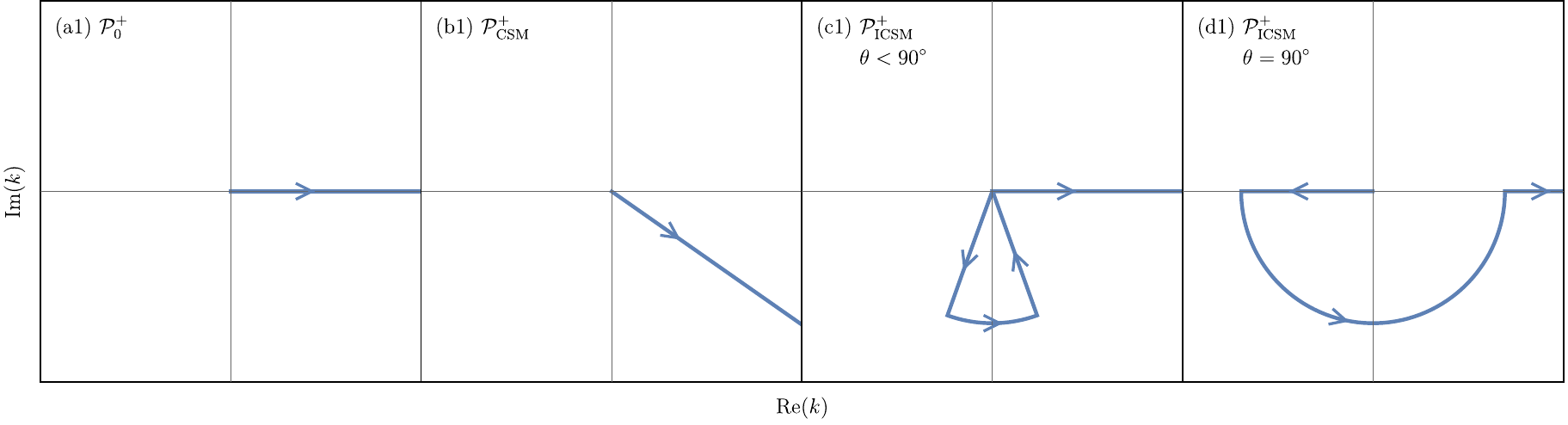}
	\end{minipage}
	}
	\subfigure
	{
	\begin{minipage}[c]{0.95\linewidth}
		\centering
		\includegraphics[width=1.\linewidth]{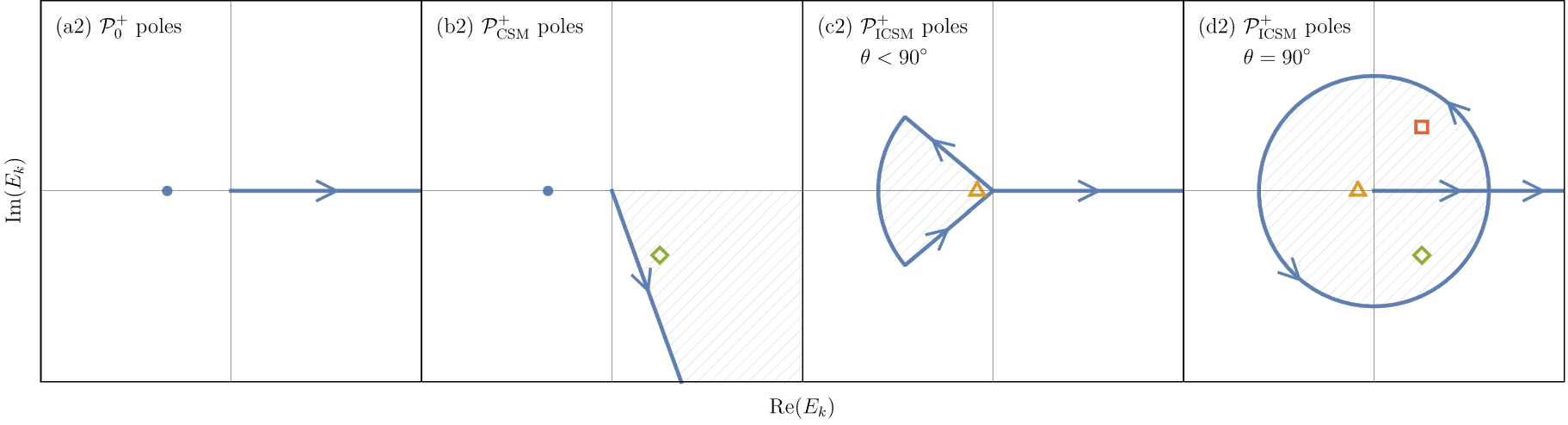}
	\end{minipage}
	}
	\subfigure
	{
	\begin{minipage}[c]{0.95\linewidth}
		\centering
		\includegraphics[width=1.\linewidth]{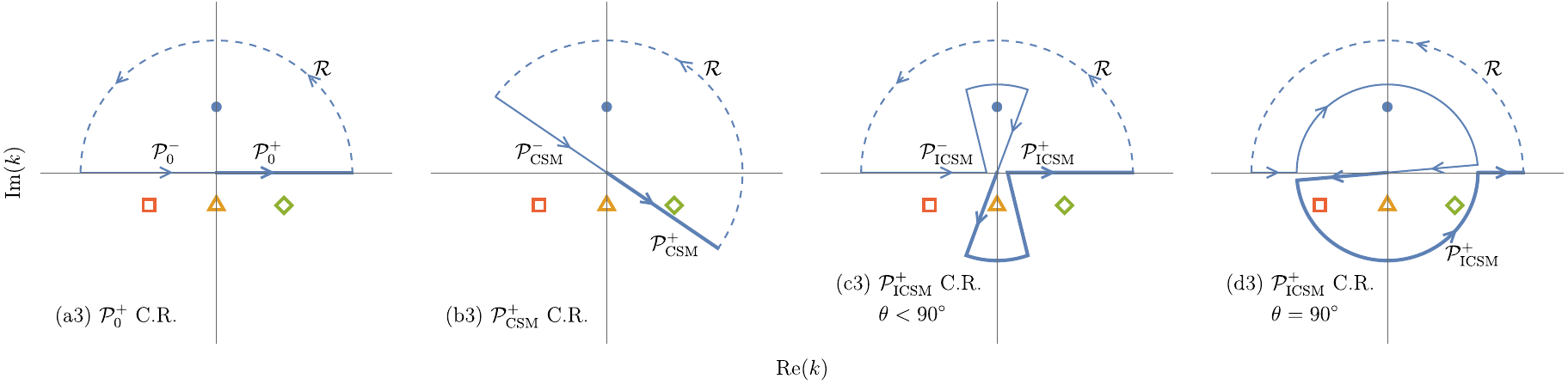}
	\end{minipage}
	}
	\caption{Various contours $\mathcal{P}^+$ in the momentum plane (a1)-(d1), the corresponding spectra in the energy plane (a2)-(d2), and the corresponding contours  $\mathcal{C}=\mathcal{P}^- +\mathcal{P}^+ +\mathcal{R}$ in the proof of the completeness relations (a3)-(d3), where $\mathcal{P}^-$ represents the path symmetric to $\mathcal{P}^+$ with respect to the origin and $\mathcal{R}$ represents the semicircular path at infinity of the upper half-plane. In the middle row, the eigenstates (including continuum states and discrete states) that could be obtained from the contours in the upper row are presented. The shaded region represents the RS-II. The solid and hollow markers represent the poles on the RS-I and -II, respectively. In the lower row, all possible poles are presented, but only the poles inside the closed contours can be obtained from the (I)CSM. In order to show the contours clearly in (c3) and (d3), we separate the six-line joints and four-line joints into several two-line joints only for illustration (not for calculation).}
	\label{fig:4_kinds_of_int_path}
\end{figure*}

The properties of the poles of the bound states, virtual states, resonances, and antiresonances can be extracted from the $T$ matrix through the LSE. For simplicity, we use the single-channel equation as an example. In the partial-wave formalism, we assume that the partial-wave potential $V_l$ does not depend on the energy explicitly. The nonrelativistic partial-wave LSE reads
\begin{equation}\label{eq:lseq}
	T_l\left(p^{\prime}, p; k\right)=V_l\left(p^{\prime}, p\right)+\int_0^{\infty} \frac{q^2 d q}{(2 \pi)^3} \frac{V_l\left(p^{\prime}, q \right) T_l\left(q, p;k \right)}{\frac{k^2}{2\mu} -\frac{q^2}{2 \mu}+i \epsilon},
\end{equation}
where $\mu$ is the reduced mass. The $p'$ and $p$ are the momentum magnitudes of the final state and initial state, respectively, in the center-of-mass frame. $E_k={k^2 \over 2\mu }$ is the scattering energy. We can obtain the on-shell $T$ matrix by taking $p'=p=k$. \clabel[momentumdependence]{The potential $V_l(p_1,p_2)$ has an explicit momentum dependence and rapidly approaches zero at large momenta to ensure the finiteness of the integral over $q$ in Eq.\eqref{eq:lseq}.} When employing the LSE to determine the pole positions, one usually needs to discretize Eq.~\eqref{eq:lseq}:
\begin{equation}\label{eq:dis_LSEQ}
    \mathbb{T}(E)=\mathbb{V}+\mathbb{V} \mathbb{G}(E) \mathbb{T}(E),
\end{equation}
where $\mathbb{T}$, $\mathbb{V}$, and $\mathbb{G}$ are matrices obtained through momentum discretization. The pole positions in Eqs.~\eqref{eq:lseq} and~\eqref{eq:dis_LSEQ} can be found as the roots of the Fredholm determinant.
\begin{equation}\label{eq:fredholm_det}
    \mathrm{Det}\left[\mathbf{1}-\mathbb{V}\mathbb{G}(E)\right]=0.
\end{equation}
In principle, one can adopt some root-finding algorithms or optimization algorithms in the entire complex $E$ plane to find the zeros of the Fredholm determinant. However, the processes are usually iterative and time consuming.

The bound state poles can be obtained directly by expanding the Hamiltonian with some square-integrable bases. For resonances, a similar calculation can be performed using the CSM proposed by Aguilar, Balslev, and Combes~\cite{Aguilar:1971ve,Balslev:1971vb}. We start from the Schr\"odinger equation in momentum space:
\begin{equation}\label{eq:p_sch_eq}
\begin{gathered}
\frac{p^2}{2 \mu} \phi_l(p)+\int_{0}^{+\infty} \frac{p'^2 d p^{\prime}}{(2 \pi)^3} V_{l}\left(p, p^{\prime}\right) \phi_{l}\left(p^{\prime}\right)=E \phi_l(p).
\end{gathered}
\end{equation}
$p'$ in Eq.~\eqref{eq:p_sch_eq} is real; namely, the integral contour $\mathcal{P}_0^+$ is along the real axis, as illustrated in Fig.~\ref{fig:4_kinds_of_int_path} (a1). Solving Eq.~\eqref{eq:p_sch_eq} may yield a sequence of discrete bound states and continuous scattering states, as shown in Fig.~\ref{fig:4_kinds_of_int_path} (a2). These states form a complete set and satisfy the completeness relation:
\begin{equation}\label{eq:CR1}
\begin{aligned}
\mathbf{1} & =\sum_b\left|\chi_b\right\rangle\left\langle\chi_b\right|+\int_0^{+\infty} d E_k\left| \chi_{E_k}\right\rangle\left\langle\chi_{E_k}\right|, \\
& =\sum_b\left|\chi_b\right\rangle\left\langle\chi_b\right|+\int_0^{+\infty} d k \left(\left| \chi_k\right\rangle\left\langle\chi_k\right|+\left| \chi_{-k}\right\rangle\left\langle\chi_{-k}\right|\right),\\ 
& =\sum_b\left|\chi_b\right\rangle\left\langle\chi_b\right|+\int_{-\infty}^{+\infty} d k\left| \chi_k\right\rangle\left\langle\chi_k\right|,
\end{aligned}
\end{equation}
where $\chi_b$ and $\chi_{E_k}$ are the bound states and continuum states on the RS-I, respectively. $\chi_k$  and $\chi_{-k}$ are continuum states on the real $k$ axis. The detailed mathematical proof of Eq.~\eqref{eq:CR1} can be found in Ref.~\cite{Newton:1960nws}. In the proof of the completeness relation, the Green's function is integrated along the contour $\mathcal{C}_0=\mathcal{P}_0^- +\mathcal{P}_0^+ +\mathcal{R}$ presented in Fig.~\ref{fig:4_kinds_of_int_path} (a3), where $\mathcal{P}_0^-$ represents the path symmetric to $\mathcal{P}_0^+$ with respect to the origin and $\mathcal{R}$ represents the semicircular path at infinity of the upper half-plane. The discrete eigenvalues (corresponding to the poles of the $T$ matrix) enclosed by $\mathcal{C}_0$ are accessible  by Eq.~\eqref{eq:p_sch_eq}.

Within the framework of the CSM, a complex scaling rotation $(p,p')\to (\tilde{p},\tilde{p}')= (pe^{-i\theta},p'e^{-i\theta}) $ is introduced to Eq.~\eqref{eq:p_sch_eq},
\begin{equation}\label{eq:csmwf}
	\frac{\tilde p^2}{2 \mu} \phi_l(\tilde p)+\int_{\mathcal{P^+_{\mathrm{CSM}}}} \frac{\tilde p^{\prime2} d \tilde p^{\prime}}{(2 \pi)^3}  V_{l}\left(\tilde p, \tilde p^{\prime} \right) \phi_{l}\left(\tilde p^{\prime}\right)=E \phi_l(\tilde p).
\end{equation}
The CSM performs a integral along the contour $\mathcal{P}^+_{\mathrm{CSM}}$, as depicted in Fig.~\ref{fig:4_kinds_of_int_path} (b1). It is important to note that the definitions of the bra and ket eigenstates differ from the conventional case here and in the following discussions. The inner products are defined using the c product, and the bra(ket)eigenstates are normalized accordingly\cite{Romo:1968tcz}:
\begin{equation}
(\chi_n | \chi_m)=\int_{\mathcal{P}_{\mathrm{CSM}}^+} \frac{k^2d k}{(2 \pi)^3} \chi_n(k)\chi_m(k)=\delta_{m,n}.
\end{equation}
With the new contour $\mathcal{P}^+_{\mathrm{CSM}}$, the modified completeness relation reads
\begin{equation}
\begin{aligned}
    \mathbf{1}=\sum_{n \in \mathrm{C}} \left|\chi_n\right)\left(\chi_n\right|+\int_{\mathcal{P}^+_{\mathrm{CSM}}+\mathcal{P}^-_{\mathrm{CSM}}} dk \left| \chi_k\right)\left(\chi_k\right|,
\end{aligned}
\end{equation}
where $\mathcal{P}^-_{\mathrm{CSM}}$ represents the path symmetric to $\mathcal{P}^+_{\mathrm{CSM}}$ with respect to the origin, and $\mathrm{C}$ refers to the region enclosed by the contour $\mathcal{C}=\mathcal{P}^-_{\mathrm{CSM}}+\mathcal{P}^+_{\mathrm{CSM}}+\mathcal{R}$, as illustrated in Fig.~\ref{fig:4_kinds_of_int_path} (b3). The mathematical proof of the completeness relation for the CSM is presented in Refs.~\cite{Giraud:2003hhs,Giraud:2004rhz}. It is notable that, within the contour illustrated in Fig.~\ref{fig:4_kinds_of_int_path} (b3), a portion of the RS-II of the energy plane is encompassed. The resonances in this region can be obtained as discrete eigenstates of Eq.~\eqref{eq:csmwf}, as depicted in Fig.~\ref{fig:4_kinds_of_int_path} (b2).

\begin{figure}[htbp]
  \includegraphics[width=.35\textwidth]{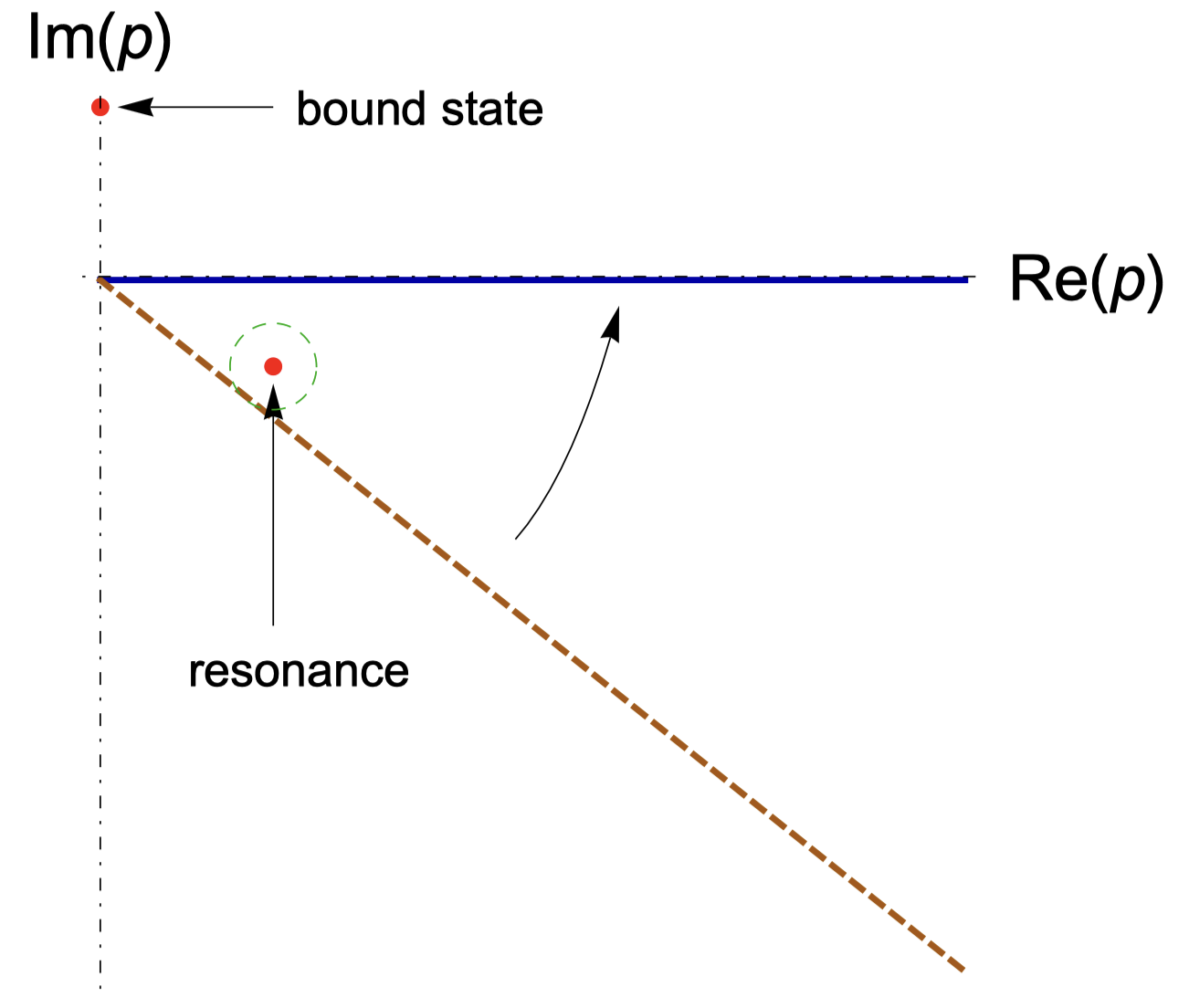}
  \caption{The differences between the poles on the RS-I and -II. The integrals along the brown dashed line and the blue solid line are equal for the bound states but different for the resonances. The correct choice of the integral contours for the resonances is the brown dashed line.}
  \label{fig:lin_integal_path}
\end{figure}

There is an alternative perspective to understand the success of the CSM~\cite{Lin:2023dbp}. A resonance solution of Eq.~\eqref{eq:csmwf} can be expressed,
\begin{equation}\label{eq:csmRwf}
	\phi_l^R(\tilde{p})=\frac{1}{E_R-\frac{\tilde{p}^2}{2 \mu}} \int_{\mathcal{P}_\mathrm{CSM}^+} \frac{\tilde{p}^{\prime 2} d \tilde p^{\prime}}{(2 \pi)^3}  V_{l}\left(\tilde{p}, \tilde p^{\prime} \right) \phi_{l}^R\left(\tilde p^{\prime} \right),
\end{equation}
where $E_R=M-i \frac{\Gamma}{2}$ is the resonance pole energy. The wave function $\phi_l^R(\tilde p)$ itself also exhibits poles at $\frac{\tilde p^2}{2\mu}=E_{\mathrm{R}}$. The pole of the wave function results in discontinuity of the integral. In Eqs.~\eqref{eq:csmwf} and \eqref{eq:csmRwf}, the integral paths above and below the pole differ by a residue of the pole, as shown in Fig.~\ref{fig:lin_integal_path}. One can get a resonance on the RS-II,
if one chooses the contour  ``below" the pole~\cite{Lin:2023dbp}.

In practical applications of the CSM, the function $V\left(\tilde p,\tilde p'\right)$ must be analytic. For instance, the complex-scaled Gaussian-type potential reads
\begin{equation*}
 e^{-(p^2+p^{\prime 2})} \rightarrow e^{(p^2+p'^2)[-\cos(2\theta)+i\sin(2\theta)]}.
\end{equation*}
For $\theta>\frac{\pi}{4}$, this potential diverges, rendering the CSM inapplicable~\cite{aoyama2006complex}. Therefore, the region of the RSs that the CSM can probe is severely limited. For a single-channel problem, it is difficult for the CSM to obtain the poles requiring large complex scaling angles, such as wide resonances and virtual states. In the coupled-channel problems, there are more RSs, exacerbating this limitation.

To improve the CSM, we choose a more flexible integral contour to solve the Schr\"odinger equation as long as the potentials remain analytic:
\begin{equation}\label{eq:aesheq}
	\frac{p^2}{2 \mu} \phi_l(p)+\int_{\mathcal{P}^+_{\mathrm{ICSM}}} \frac{p^{\prime 2} d p^{\prime}}{(2 \pi)^3} V_{l}\left(p, p^{\prime}\right) \phi_{l}\left(p^{\prime}\right)=E \phi_l(p),
\end{equation}
where both $p$ and $p'$ are complex momenta along a general contour $\mathcal{P}^+_{\mathrm{ICSM}}$, ensuring the analyticity of the potential. The extended completeness relation of Eq.~\eqref{eq:aesheq} becomes
\begin{equation}
\begin{aligned}
    \mathbf{1}=\sum_{n \in \mathrm{C}} \left|\chi_n\right)\left(\chi_n\right|+\int_{\mathcal{P}^+_{\mathrm{ICSM}}+\mathcal{P}^-_{\mathrm{ICSM}}} dk \left| \chi_k\right)\left(\chi_k\right|,
\end{aligned}
\end{equation}
where $\mathcal{P}^-_{\mathrm{ICSM}}$ represents the path symmetric to $\mathcal{P}^+_{\mathrm{ICSM}}$ with respect to the origin, and $\mathrm{C}$ refers to the region enclosed by the contour $\mathcal{C}=\mathcal{P}^-_{\mathrm{ICSM}} +\mathcal{P}^+_{\mathrm{ICSM}} +\mathcal{R}$. The mathematical proof of the extended completeness relation can be found in Refs.~\cite{Berggren:1968zz,Lind:1993zz,Giraud:2004rhz,Giraud:2003hhs}. In principle, all the poles inside the contour $\mathcal{C}$ can be obtained by solving the eigenvalue problem of the Schr\"odinger equation. One can adapt the contours to suit the specific problems.

In order to detect the virtual state poles below the threshold on the RS-II in a single-channel problem, we design the $\mathcal{P}^+_{\mathrm{ICSM}}$ as
\begin{equation}\label{eq:ICSM_path}
\begin{aligned}
		\mathcal{P}_{\mathrm{ICSM}}^+:~0 &\rightarrow \sqrt{2\mu E_0} e^{-i({\pi \over 2}+\theta)}\to \sqrt{2\mu E_0} e^{-i({\pi \over 2}-\theta)}\\
		&\rightarrow (0+\epsilon) \rightarrow +\infty.
\end{aligned}
\end{equation}
The contour $\mathcal{P}_{\mathrm{ICSM}}^+$ for $\theta < {\pi \over 2}$ is depicted in Fig.~\ref{fig:4_kinds_of_int_path} (c1). The corresponding contour of the Green's function in the proof of the completeness relation is illustrated in Fig.~\ref{fig:4_kinds_of_int_path} (c3). The poles enclosed by the contour in Fig.~\ref{fig:4_kinds_of_int_path} (c3),  including virtual states, can then be obtained as discrete eigenstates of Eq.~\eqref{eq:aesheq}, as shown in Fig.~\ref{fig:4_kinds_of_int_path} (c2). One can also choose $\theta ={\pi \over 2}$, with the corresponding contour of Schr\"odinger equation in Fig.~\ref{fig:4_kinds_of_int_path} (d1), the contour of the Green's function in Fig.~\ref{fig:4_kinds_of_int_path} (d3) and the possible spectrum in Fig.~\ref{fig:4_kinds_of_int_path} (d2).

We also explore a more intricate multichannel problem involving more RSs. We apply distinct contours for different channels to ensure that all relevant poles can be detected. Detailed discussions can be found in Sec.~\ref{sec:multichannel}. It is important to note that the choice of the contours is very flexible to enclose the target poles depending on the specific problem. The only essential requirement is the analyticity of the potential.

We use the Gauss-Legendre quadrature in pieces to perform contour integrations. We divide the contours into several smooth pieces. Several tens of Gauss points for each piece shall generate very accurate results. Thus, the (I)CSM is very stable and efficient. \clabel[systematicerro]{Similar to employing the root-finding method to solve the LSE, the only systematic error of using (I)CSM to identify poles arises from numerically solving the integral equation. The systematic errors are negligible. The numerical algorithms are not the main focus of this paper and therefore we do not explicitly present the systematic errors in the numerical results.}

\section{Single-channel case}\label{sec:single-channel}

In this section, we focus on the single-channel problems as demonstrative instances. We apply the ICSM to a separable toy potential and two realistic physical systems: the $D D^*$ potential from the lattice QCD simulation and the high-quality $NN$ Reid93 potential. We conduct a comparative analysis through the application of the LSE, CSM, and ICSM techniques. The approach to solve the LSE is similar to the technique employed in Refs.~\cite{Meng:2020cbk, Meng:2021jnw}. 

\subsection{Separable potential}
\begin{figure}[htbp]
	\centering
	\includegraphics[width=.45\textwidth]{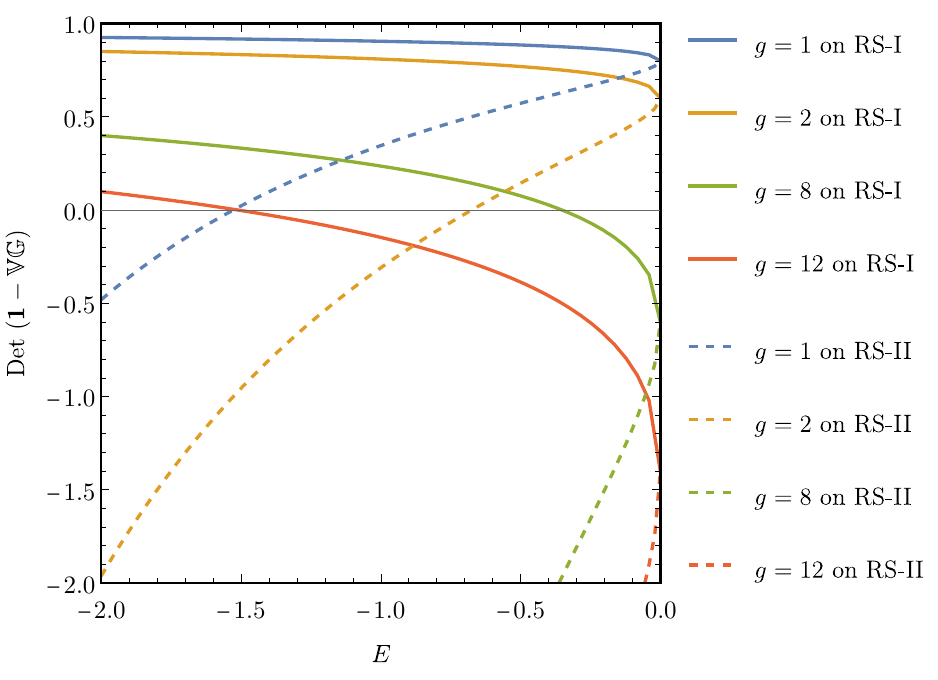}
	\caption{Variation of $\mathrm{Det}(\mathbf{1}-\mathbb{V}\mathbb{G})$ with the energy for different values of $g$. The solid (dashed) lines represent the results on RS-I (-II). The zeros of the solid (dashed) curves indicate the bound (virtual) state  poles with $g=8$ and $12$ ($g=1$ and $2$).}
	\label{fig:exp1_LS_gall}
\end{figure}

We first consider a simple single-channel separable potential:
\begin{equation}\label{eq:potential_vex1}
  V_{\mathrm{Ex.1}}(p,p')=-\frac{g}{4\pi} \frac{1000}{\left(p^2+10\right)\left(p'^2+10\right)},
\end{equation}
with a dimensionless reduced mass $\mu=1$, and the orbital angular momentum $L=0$. We solve the LSE to obtain the $\mathrm{Det}(\mathbf{1}-\mathbb{V}\mathbb{G})$ as a benchmark. In Fig.~\ref{fig:exp1_LS_gall}, we present the $\mathrm{Det}(\mathbf{1}-\mathbb{V}\mathbb{G})$ with respect to the energy on RS-I and RS-II, considering different values of $g$. The pole positions on different sheets can be identified from the zeros of the curves. As the strength of the attractive potential $g$ decreases, a bound state pole $(g=8~\text{and}~12)$ becomes a virtual state pole $(g=1~\text{and}~2)$.

\begin{figure*}[htbp]
	\centering
	\includegraphics[width=.9\linewidth]{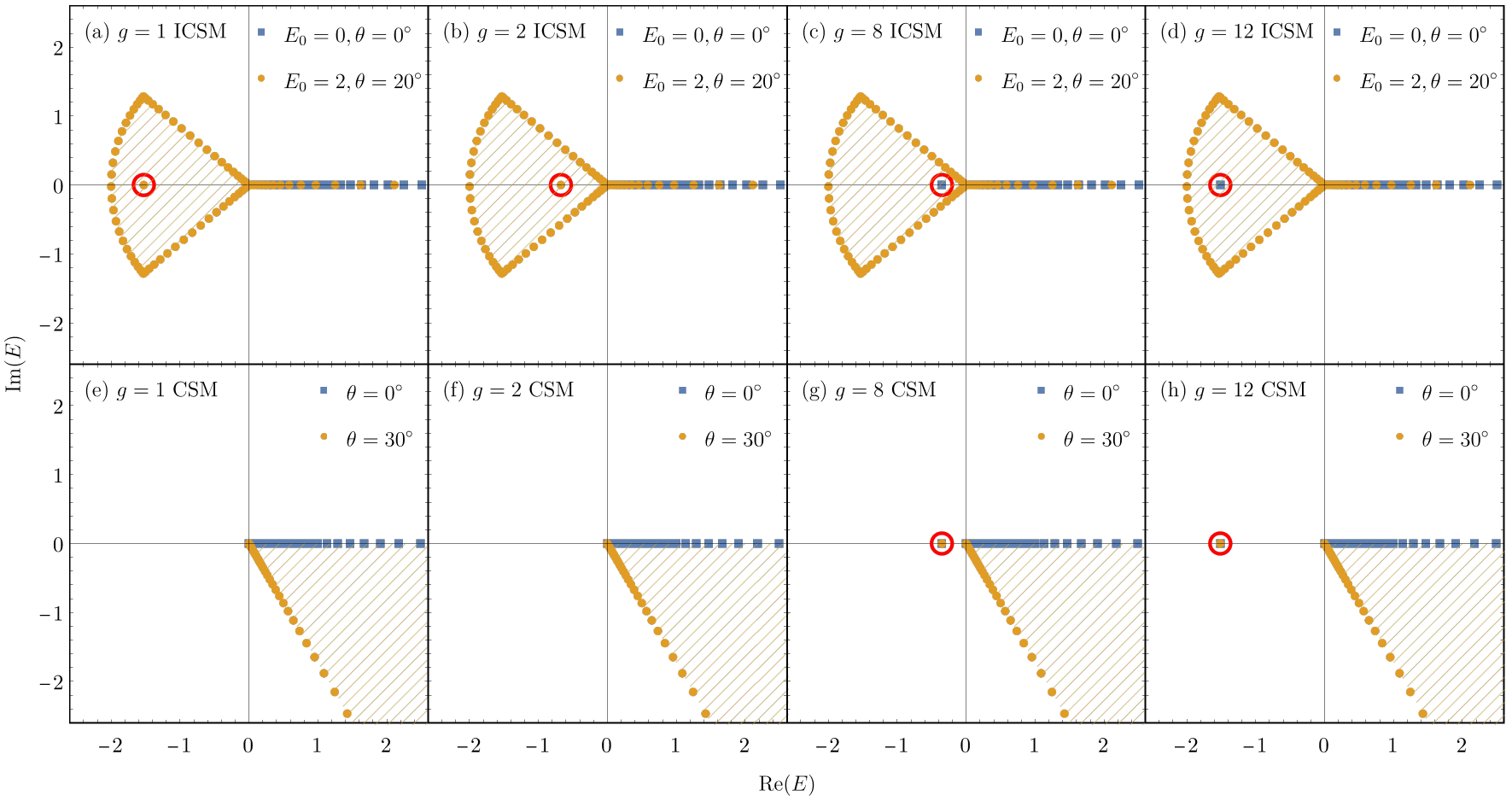}
	\caption{Complex eigenenergies obtained using the (a)-(d) ICSM and (e)-(h) CSM for $V_{\mathrm{Ex.1}}$. The locations where poles appear are marked with red circles. The yellow shaded region represents the RS-II enclosed by the yellow contour.}
	\label{fig:exp1_ICSM}
\end{figure*}

Furthermore, we employ the ICSM and CSM to solve the Schr\"odinger equation, respectively. The results are presented in Fig~\ref{fig:exp1_ICSM}. In the ICSM, we get the virtual state poles within the yellow shaded region for $g=1$ and $2$. For $g=8$ and $12$, the bound state poles are obtained from the integral path along the real axis. On the other hand, the CSM finds only the bound state poles.

The relevant numerical results are summarized in Table~\ref{tab:exp1}. In the single-channel toy models, both ICSM and LSEs yield the consistent bound state and virtual state poles, while the CSM can detect only the bound state poles.
\begin{table}[htbp]
    \centering
    \caption{The pole positions and their corresponding RSs obtained from the LSEs, ICSM, and CSM with different coupling constants $g$ in $V_{\mathrm{Ex.1}}$.}
    \label{tab:exp1}
    \begin{tabular}{l|cc|cc|cc}
    \hline\hline
         \multirow{2}{*}{~strength~} & \multicolumn{2}{c|}{LSEs} & \multicolumn{2}{c|}{ICSM} & \multicolumn{2}{c}{CSM}\\
         & $E_{\mathrm{pole}}$ & RS & $E_{\mathrm{pole}}$ & RS & $E_{\mathrm{pole}}$ & RS  \\ \hline
        $~~g=1$ & ~~~~$-1.526$~~~~ & ~~~~II~~~~ & ~~~~$-1.526$~~~~ & ~~~~II~~~~ & ~~~~None~~~~ &   \\ 
        $~~g=2$ & $-0.674$ & II & $-0.674$ & II & None &   \\ 
        $~~g=8$ & $-0.353$ & I & $-0.353$ & I & $-0.353$ & ~~~~I~~~~ \\ 
        $~~g=12$ & $-1.513$ & I & $-1.513$ & I & $-1.513$ & I \\ \hline\hline
    \end{tabular}
\end{table}

\subsection{High-quality $NN$ Reid93 potential}

\begin{figure*}[tbp]
	\centering
	\subfigure
	{
	\begin{minipage}[c]{0.558\linewidth}
		\centering
		\includegraphics[width=1\linewidth]{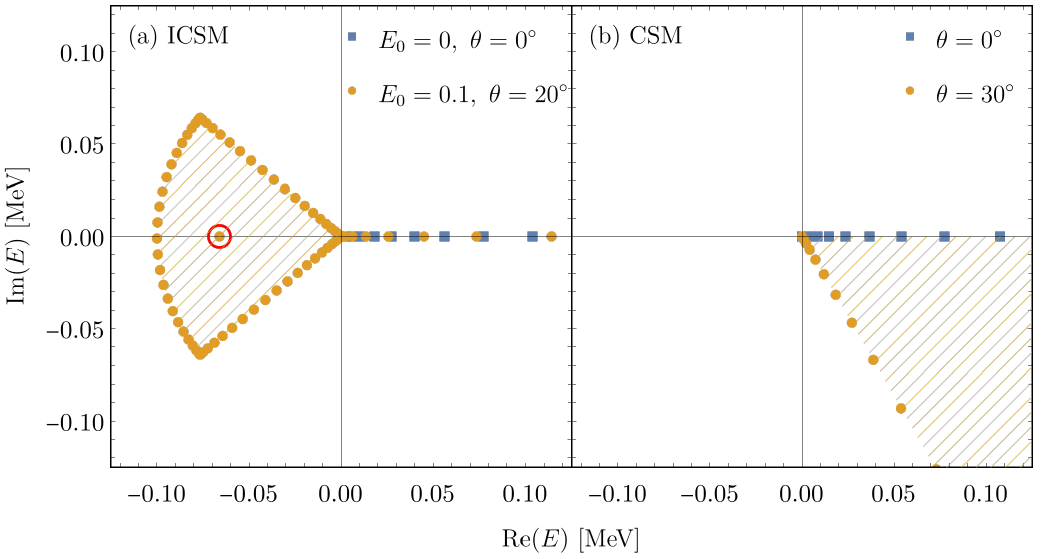}
	\end{minipage}
	}
	\subfigure
	{

	\begin{minipage}[c]{0.314\linewidth}
		\centering
		\includegraphics[width=1\linewidth]{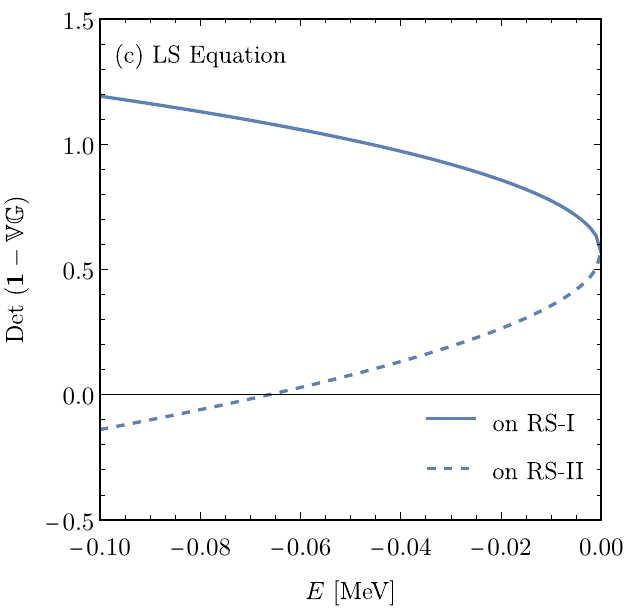}
	\end{minipage}
	}
	\caption{Complex eigenenergies obtained using the (a) ICSM, (b) CSM, and the variation of $\mathrm{Det}(\mathbf{1}-\mathbb{V}\mathbb{G})$ with the energy obtained using the (c) LSE for $V_{np}(^1S_0)$. The notations in the figure are the same as those in Figs.~\ref{fig:exp1_LS_gall} and~\ref{fig:exp1_ICSM}.}
	\label{fig:Reid93}
\end{figure*}
In nuclear physics, the virtual state of $np~(^1S_0)$ is of great interest \cite{Adhikari:1982zzb,Carlson:1997zzb,Ren:2017yvw,Borzakov:2021moq}. We test the ICSM using the high-quality $NN$ potential called Reid93 in Ref.~\cite{Stoks:1994wp}, which accurately describes the experimental data of the $np$ scattering. 
\begin{eqnarray}
	V_{n p}\left({ }^1 S_0\right)&=&\sum_{i=3}^6B_{1,i} Y(i)\nonumber\\ 
 & &+f_{\pi}^2\left(-V_{\mathrm{OBE},\pi^0}+2V_{\mathrm{OBE},\pi^\pm}\right),\\
	Y(n,q)&=&n \bar{m}_{\pi}\phi^0(n \bar{m}_{\pi}, \bar \Lambda_\pi, q),\\
 V_{\mathrm{OBE}}(m,\Lambda,q)&=&\left(\frac{m}{m_{\pi^\pm}}\right)^2\frac{m}{3}\phi^1(m,\Lambda,q)(\bm \sigma_1\cdot\bm \sigma_2),\quad\\
	\phi^n\left(m,\Lambda,q\right)&=&\frac{4\pi\left(-q^2\right)^n}{m^{2n+1}\left(q^2+m^2\right)}\left(\frac{\Lambda^2-m^2}{\Lambda^2+q^2}\right)^2.\qquad
\end{eqnarray}
All the parameters can be found in~\cite{Stoks:1994wp}. We employ the LSE, ICSM, and CSM to investigate the pole positions with this potential, and the results are shown in Fig.~\ref{fig:Reid93}. The LSE and ICSM yield a virtual state pole at $E_{np,v}=-0.066~\mathrm{MeV}$, while the CSM fails to obtain the virtual state.

\subsection{HAL QCD potential of $D D^*$ system}
\begin{figure*}[tbp]
	\centering
	\subfigure
	{
	\begin{minipage}[c]{0.58\linewidth}
		\centering
		\includegraphics[width=1\linewidth]{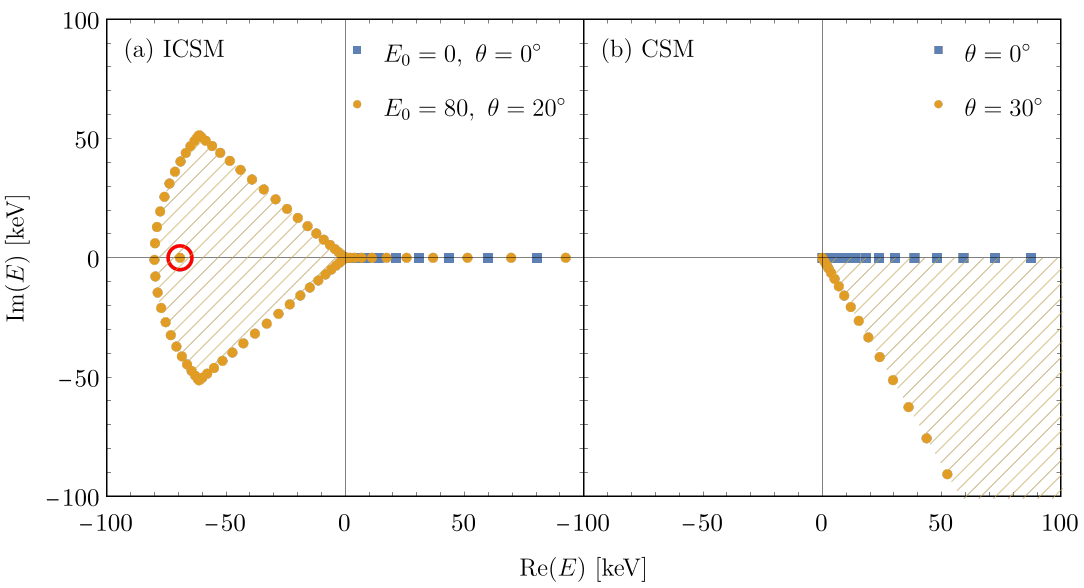}
	\end{minipage}
	}
	\subfigure
	{
	\begin{minipage}[c]{0.314\linewidth}
		\centering
		\includegraphics[width=1\linewidth]{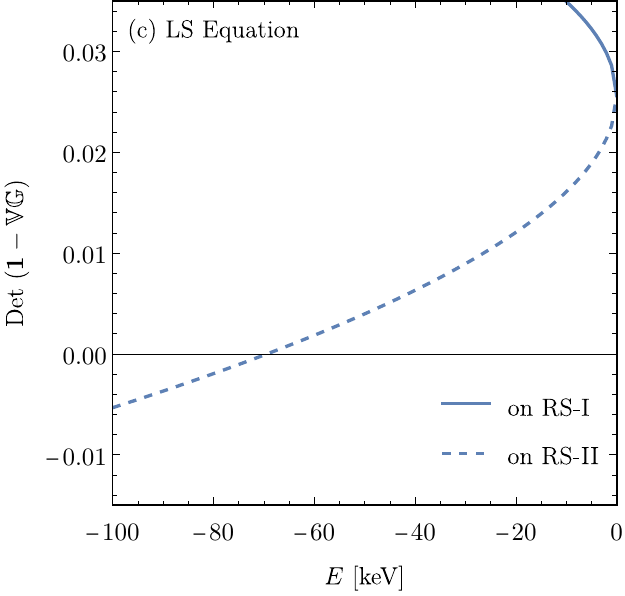}
	\end{minipage}
	}
	\caption{Complex eigenenergies obtained using (a) ICSM, (b) ICSM, and the variation of $\mathrm{Det}(\mathbf{1}-\mathbb{V}\mathbb{G})$ with the energy obtained using the (c) LSE for $V_{DD^*}$. The notations are the same as those in Figs.~\ref{fig:exp1_LS_gall} and~\ref{fig:exp1_ICSM}.}
	\label{fig:HALQCD}
\end{figure*}

Recently, the $T_{cc}^+$ state was observed by the LHCb Collaboration~\cite{LHCb:2021vvq,LHCb:2021auc}, which has sparked heated discussions on its nature~\cite{Li:2012ss,Meng:2021jnw,Deng:2021gnb,Du:2021zzh,Ortega:2022efc,Cheng:2022qcm,Wang:2022jop,Wang:2023ovj,Du:2023hlu,Lyu:2023xro}. Especially, the $T_{cc}^+$ state was investigated in the molecular scheme via the lattice QCD simulation in a potential formalism~\cite{Lyu:2023xro}, namely the HAL QCD method. The isoscalar and S-wave channel $D D^*$ potential is extracted in discrete space points. Subsequently, an correlated fit for the potential is conducted, employing a phenomenological four-range Gaussian form:
\begin{equation}
  V_{DD^*}(r)=\sum_{i=1}^{4} a_i e^{-\left(r / b_i\right)^2},
\end{equation}
with the parameters
\begin{equation}
  \begin{gathered}
  	\left(a_1, a_2, a_3,a_4\right)=(-269,-121,-81,-23)~\mathrm{MeV},\\
  	\left(b_1, b_2, b_3,b_4\right)=(0.14,0.27,0.52,0.97)~\mathrm{fm},\\
  	\left(m_{D},m_{D^*}\right)=(1878.2, 2018.1)~\mathrm{MeV}.
  \end{gathered}
\end{equation}
The authors employ an approximation method based on the effective range expansion and obtain a virtual state pole with $E_{DD^*,v}^{\mathrm{ERE}}\approx -59\mathrm{keV}$.

In order to perform the ICSM calculation, we transform this potential in momentum space:
\begin{equation}
  V_{DD^*}\left(q\right)=\sum_{i=1}^{4} a_i\left(\pi b_i^2\right)^{3 / 2} e^{-\frac{b_i^2}{4}q^2},
\end{equation}
where $\bm q=\bm{p}-\bm{p}^{\prime}$. We employ the LSE, ICSM, and CSM for $V_{DD^*}$ to search for the poles, and the results are depicted in Fig.~\ref{fig:HALQCD}. Similar to the previous cases, the LSE and ICSM yield a virtual state pole at $E^{\mathrm{ICSM}
}_{DD^*,v}=-69~\mathrm{keV}$, while the CSM fails to obtain any poles. 

Our results qualitatively agree with the results obtained through the effective range expansion \cite{Lyu:2023xro}. However, we avoid using approximations in our calculations. For the poles very close to the threshold, the ICSM shall be a superior approach to precisely determine which RS the pole lies on. 

\section{multichannel case}\label{sec:multichannel}

In this section, we consider the multichannel case with a separable potential:
\begin{equation}
V_{\mathrm{Ex.2}}(p,p')=-\frac{1}{4\pi}\left(\begin{array}{cc}
g_{1} & g_c \\
g_c & g_{2}
\end{array}\right)\frac{1000}{\left(p^2+10\right)\left(p'^2+10\right)}.
\end{equation}
Here, the dimensionless reduced masses for the two channels are taken as $\mu_1=\mu_2=1$, the difference between the two thresholds is set to $m_{\mathrm{th},2}-m_{\mathrm{th},1}=2$, and the orbital angular momentum is $L=0$. We select two sets of the coupling constants, namely, $(g_1,g_2,g_c)=(2,8,1.5)$ and $(g_1,g_2,g_c)=(8,2,1.5)$. For convenience, we label the four RSs as $(\mathrm{RS}_1,\mathrm{RS}_2)$, with $\mathrm{RS}_i$=I and II representing the $i$ th channel with the positive and negative imaginary parts of the scattering momentum $k_i=\sqrt{2\mu_i(E-m_{{\mathrm{th},i}})}$, respectively.

\begin{figure*}[tbp]
	\centering

	\subfigure[~~$g_1=2,~g_2=8,~g_c=1.5$]
	{
	\begin{minipage}[c]{1\linewidth}
		\centering
		\includegraphics[width=1\linewidth]{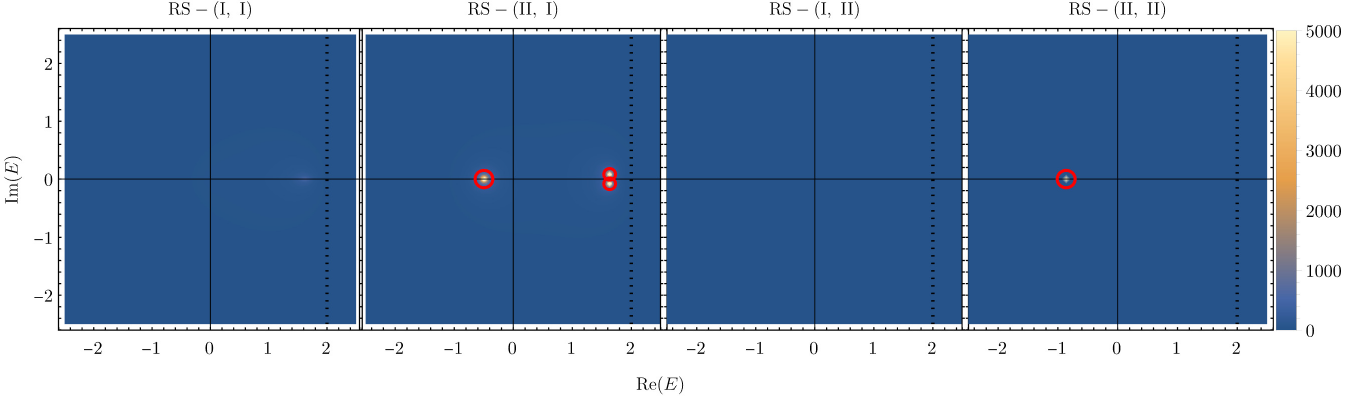}
	\end{minipage}
	}
	\subfigure[~~$g_1=8,~g_2=2,~g_c=1.5$]
	{
	\begin{minipage}[c]{1\linewidth}
		\centering
		\includegraphics[width=1.\linewidth]{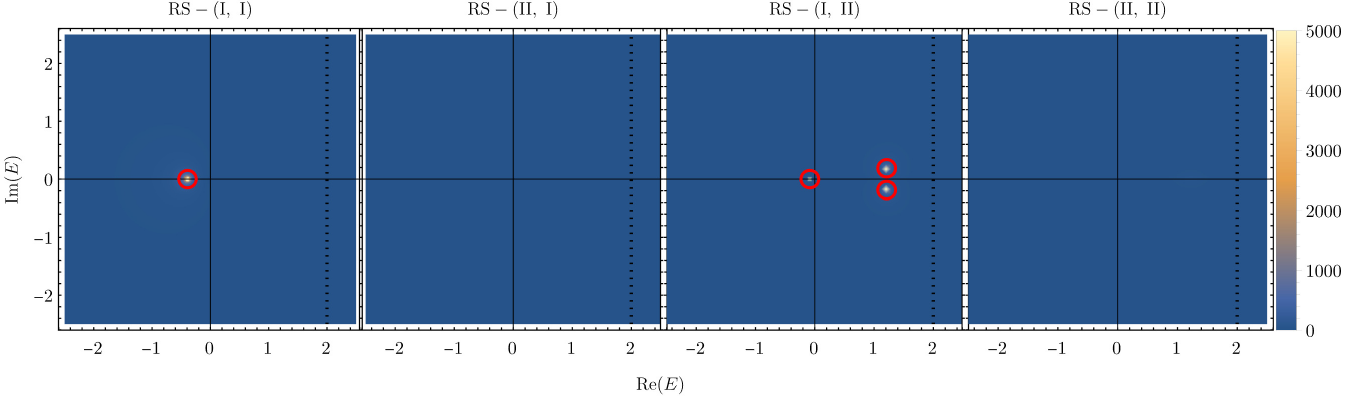}
	\end{minipage}
	}
	\caption{$\left|\mathrm{Det}(\mathbf{1}-\mathbb{V}\mathbb{G})\right|^{-2}$ on different energy RSs for the two sets of different coupling constants: (a) $(g_1,~g_2,~g_c)=(2,~8,~1.5)$, and (b) $(g_1,~g_2,~g_c)=(8,~2,~1.5)$ in $V_{\mathrm{Ex.2}}$. The locations where the poles appear are marked with red circles.}
	\label{fig:exp2_lseq}
\end{figure*}

The $\left|\mathrm{Det}(\mathbf{1}-\mathbb{V}\mathbb{G})\right|^{-2}$ from LSEs on different RSs for the two sets of the coupling constants are shown in Fig.~\ref{fig:exp2_lseq}. The poles obtained with the different coupling constants lie on distinct RSs. For $(g_1,g_2,g_c)=(2,8,1.5)$, the poles lie on RS-$\mathrm{(II,I)}$ and -$\mathrm{(II,II)}$, while for $(g_1,g_2,g_c)=(8,2,1.5)$, the poles are located on RS-$\mathrm{(I,I)}$ and -$\mathrm{(I,II)}$.

\begin{figure*}[tbp]
	\centering
	\includegraphics[width=.9\linewidth]{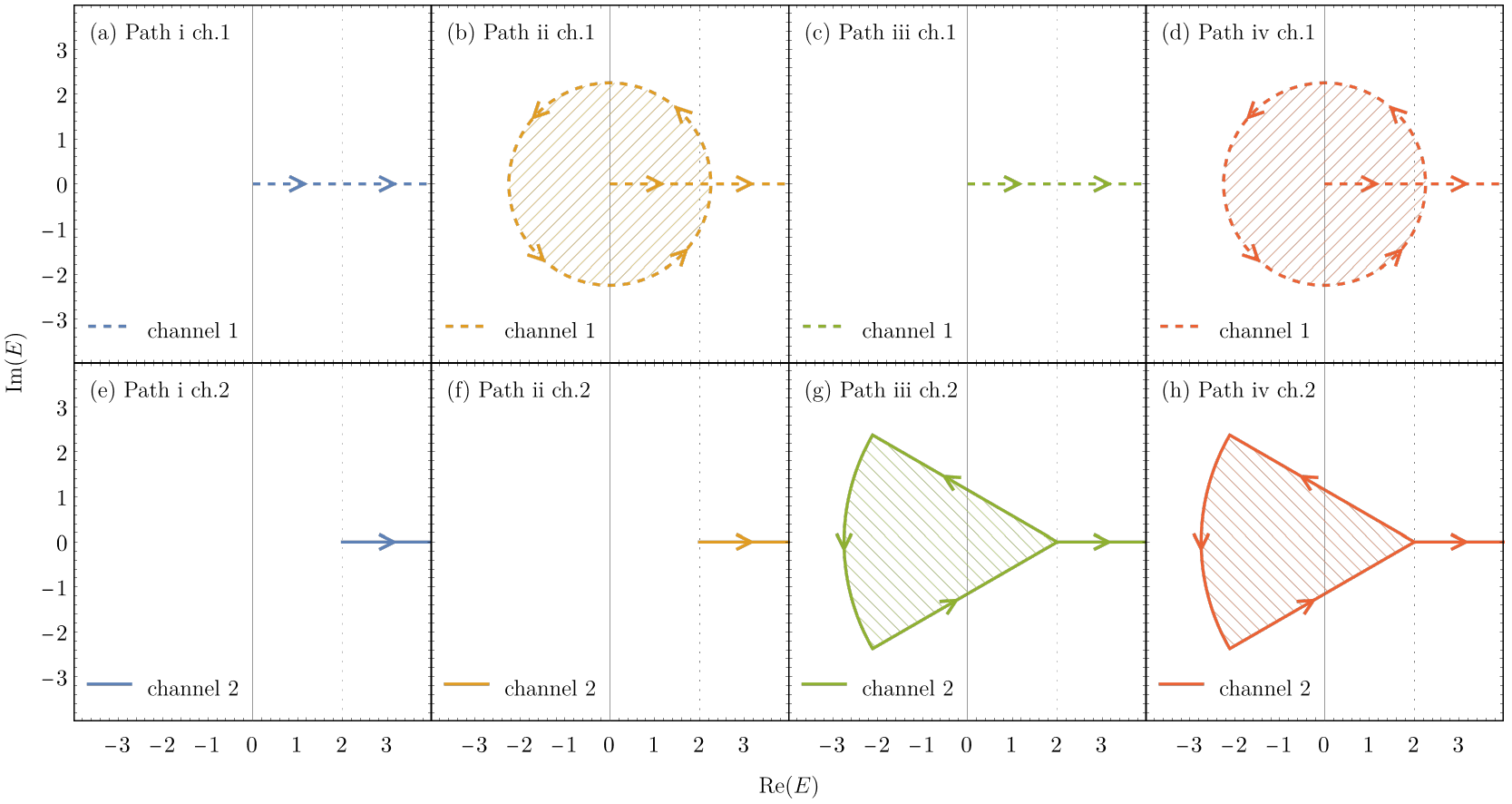}
	\caption{The four contours determined by Eq.~\eqref{eq:ex2_ICSM_path} in the energy plane. The dashed (solid) line represents the contour of channel 1 (2). The shaded region represents the RS-II enclosed by the contours for each channel. The arrow represents the direction of the contour.}
	\label{fig:exp2_ICSMpath}
\end{figure*}

\begin{figure*}[htbp]
	\centering
	\includegraphics[width=.875\linewidth]{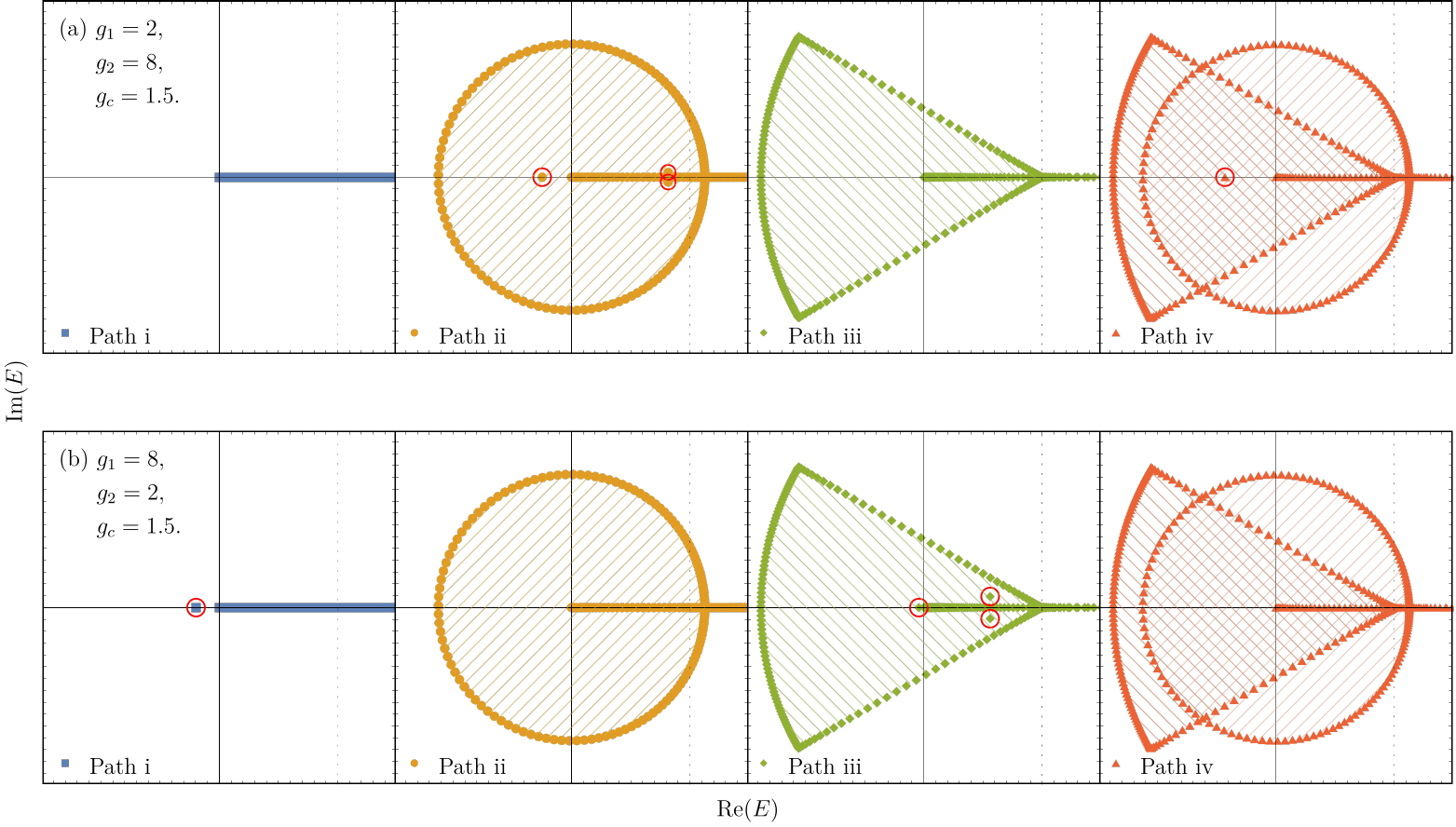}
	\caption{Complex eigenenergies obtained using the ICSM for the two sets of the coupling constants: (a) $(g_1,g_2,g_c)=(2,8,1.5)$, and (b) $(g_1,g_2,g_c)=(8,2,1.5)$ in $V_{\mathrm{Ex.2}}$.  The contours (i)-(iv) of the ICSM are shown in Fig.~\ref{fig:exp2_ICSMpath}. The bottom-left diagonal shading represents RS-(II,I), while the bottom-right diagonal shading represents RS-(I,II), and the overlapping region of them corresponds to RS-(II,II). The locations where the poles appear are marked with red circles. }
	\label{fig:exp2_ICSM}
\end{figure*}

\begin{figure}[htbp]
	\centering
	\includegraphics[width=0.475\textwidth]{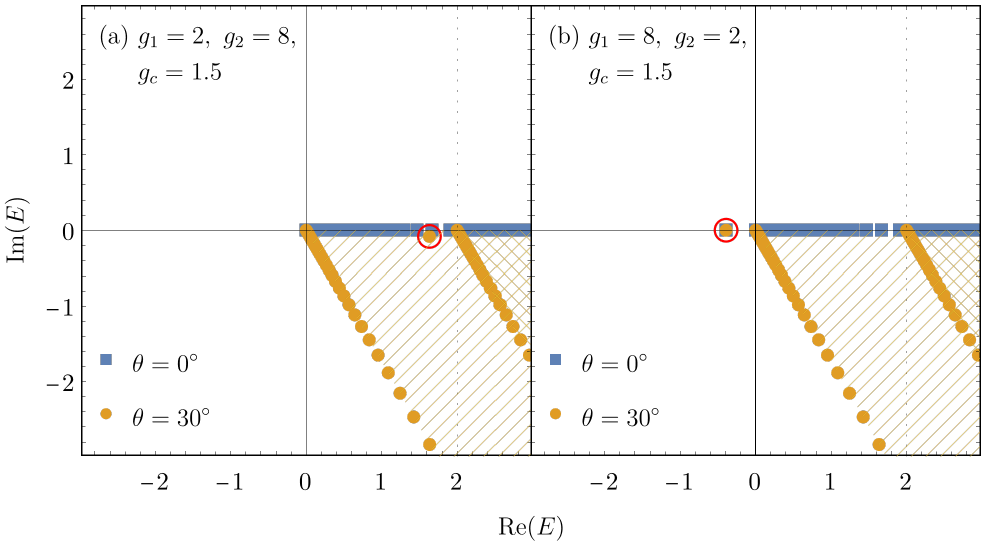}
	\caption{Complex eigenenergies obtained using the CSM for the two sets of the coupling constants: (a) $(g_1,g_2,g_c)=(2,8,1.5)$, and (b) $(g_1,g_2,g_c)=(8,2,1.5)$ in $V_{\mathrm{Ex.2}}$. The locations where the poles appear are marked with red circles. The bottom-left diagonal shading represents RS-(II,I), while the bottom-right diagonal shading represents RS-(I,II), and the overlapping region of them corresponds to RS-(II,II). }
	\label{fig:exp2_CSM}
\end{figure}

\begin{table*}[htbp]
    \centering
    \caption{The pole energies and their corresponding RSs are obtained using the LSEs, ICSM, and CSM for different values of $(g_1,g_2,g_c)$ in $V_{\mathrm{Ex.2}}$. The LSEs and ICSM yield the same results, while the CSM provides only the bound state and the quasibound state.}
    \label{tab:exp2}
    \begin{tabular}{l|cc|cc|cc}
    \hline\hline
         \multirow{2}{*}{~$(g_1,~g_2,~g_c)$~} & \multicolumn{2}{c|}{LSEs} & \multicolumn{2}{c|}{ICSM} & \multicolumn{2}{c}{CSM}\\
         & $E_{\mathrm{pole}}$ & RS & $E_{\mathrm{pole}}$ & RS & $E_{\mathrm{pole}}$ & RS  \\ \hline
         \multirow{4}{*}{~$(2,~8,~1.5)$~} & $-0.496$ & $\mathrm{(II,~I)}$ &  $-0.496$ & $\mathrm{(II,~I)}$ & None & \\
         & ~~$1.633-0.082 i$~~ & ~~$\mathrm{(II,~I)}$~~ & ~~$1.633-0.082 i$~~ & ~~$\mathrm{(II,~I)}$~~ & ~~$1.633-0.082 i$~~ & ~~$\mathrm{(II,~I)}$~~ \\
         & $1.633+0.082 i$ & $\mathrm{(II,~I)}$ & $1.633+0.082 i$ & $\mathrm{(II,~I)}$ & None & \\
         & $-0.862$ & $\mathrm{(II,~II)}$ &  $-0.862$ & $\mathrm{(II,~II)}$ & None & \\
         \hline
         \multirow{4}{*}{~$(8,~2,~1.5)$~} & $-0.392$ & $\mathrm{(I,~I)}$ &  $-0.392$ & $\mathrm{(I,~I)}$ & $-0.392$ & $\mathrm{(I,~I)}$ \\
         & ~~$-0.083$~~ & ~~$\mathrm{(I,~II)}$~~ & ~~$-0.083$~~ & ~~$\mathrm{(I,~II)}$~~ & ~~None~~ & \\
         & $1.213-0.187 i$ & $\mathrm{(I,~II)}$ & $1.213-0.187 i$ & $\mathrm{(I,~II)}$ & None & \\
         & $1.213+0.187 i$ & $\mathrm{(I,~II)}$ & $1.213+0.187 i$ & $\mathrm{(I,~II)}$ & None & \\
         \hline\hline
         \end{tabular}
\end{table*}

Then, we employ the ICSM and CSM for numerical calculations. In the ICSM, we apply distinct contours for the two channels (as illustrated in Fig.~\ref{fig:exp2_ICSMpath}) to ensure that all relevant poles can be detected. Each contour is defined by the parameters in Eq.~\eqref{eq:ICSM_path}, given as follows:
\begin{equation}\label{eq:ex2_ICSM_path}
\begin{array}{lll}
\text{Path i :}\quad & E_{0,1}=0, & \theta_1=0^\circ,\quad \\
	 & E_{0,2}=0, & \theta_2=0^\circ .\\
\text{Path ii:}\quad & E_{0,1}=2.25, & \theta_1=90^\circ,\quad \\
	 & E_{0,2}=0, & \theta_2=0^\circ .\\
\text{Path iii:}\quad & E_{0,1}=0, & \theta_1=0^\circ,\quad \\
	 & E_{0,2}=4.75, & \theta_2=15^\circ .\\
\text{Path iv:}\quad & E_{0,1}=2.25, & \theta_1=90^\circ,\quad \\
	 & E_{0,2}=4.75, & \theta_2=15^\circ .\\
\end{array}
\end{equation}
The complex eigenenergies are depicted in Figs.~\ref{fig:exp2_ICSM} and~\ref{fig:exp2_CSM}. In addition to the continuum solutions along the integral contour, one can identify some discrete solutions, which are the poles of the $T$ matrix.  The numerical results obtained through the three methods are summarized in Table~\ref{tab:exp2}. Again, in this multichannel problem, the LSE and ICSM yield the same results, while the CSM can provide only a subset of the poles, namely, the bound state on the RS-$\mathrm{(I,I)}$ and the quasibound state on the RS-$\mathrm{(II,I)}$.

\begin{figure}[tbp]
	\centering
	\includegraphics[width=1\linewidth]{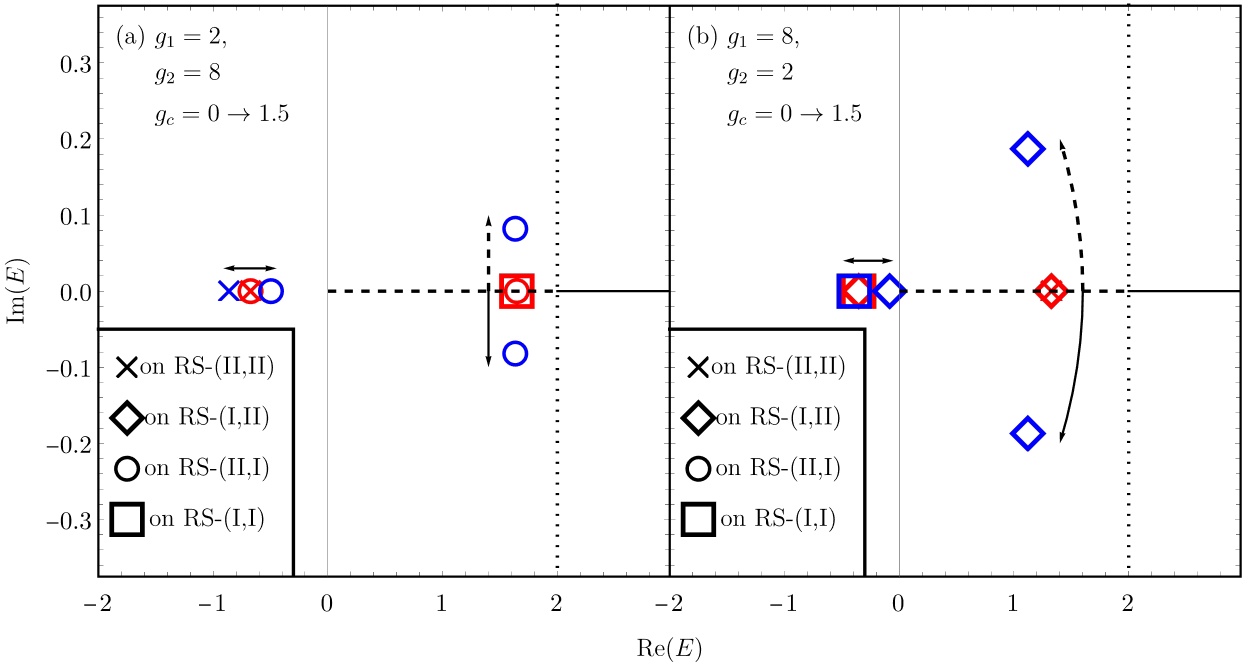}
	\caption{The pole trajectories of the coupled-channel problem as $g_c$ is changed from 0 to 1.5. The red (blue) markers label the poles with $g=0$ ($g=1.5$). The arrows indicate the trajectory of the pole shifting caused by the coupled-channel effects. The dashed arrows indicate the pole trajectory intersects a branch cut during the shift.} 
	\label{fig:exp2_pole_trace}
\end{figure}

In Fig.~\ref{fig:exp2_pole_trace}, we demonstrate the mechanism of forming the pole pattern in the double-channel example. We first neglect the coupled-channel effect by taking $g_c=0$. We can treat the two channels as two single-channel problems. For the case $(g_1,~g_2)=(2,~8)$, $g_1=2$ permits a virtual state on the RS-II of the first channel. When we take the second channel into consideration,  the virtual state pole of the first channel will not be affected since $g_c=0$. The single-channel pole splits into two poles at the same position of two different RSs, RS-(II, I) and RS-(II, II). When $g_c\neq 0$, the pole positions will shift. Similarly, for $g_2=8$, a bound state is permitted on RS-I of the second channel. In the case of $g_c=0$, this bound state pole splits into two poles on RS-(II, I) and RS-(I, I). With a nonzero $g_c$, the pole positions shift, and the RS-(I, I) pole crosses the branch cut of the first channel, transitioning to RS-(II, I), as shown in Fig.~\ref{fig:exp2_pole_trace} (a). This pair of poles on RS-(II, I) is also referred to as ``quasibound states."

The scenario of $(g_1,~g_2)=(8,2)$ is remarkably analogous, as depicted in Fig.\ref{fig:exp2_pole_trace} (b). In this scenario, the initial virtual state of the second channel transforms into a pair of poles located on RS-(I,II). These poles are termed as ``inelastic-virtual states."

It is worth noting that the separable potential represents just a special case. For a multitude of realistic potentials, solving the LSE analytically is not possible, thus necessitating numerical approaches. As emphasized previously, when dealing with the multichannel problems, the multiple RSs lead to more complicated Fredholm determinant problems, which typically involve iterative and very time-consuming numerical procedures. On the other hand, the ICSM is a simpler approach and can be directly extended to multichannel scenarios with non-separable potentials. Therefore the ICSM is considerably more efficient in determining the pole positions. 

\section{SUMMARY}\label{sec:summary}

In this work, we have proposed a strategy to improve the complex scaling method, which overcomes the limitations of the conventional CSM. \clabel[ICSMvsCSMp1]{The ICSM represents a more general case of the conventional CSM. In the conventional CSM, the transformation of the integral path is restricted to $p \rightarrow p e^{-i \theta}$, while in the ICSM, we extend this transformation to a more general and flexible form.} We perform the calculation in momentum space and use more flexible contours to solve the Schr\"odinger equation rather than use the complex-scaled contours. This method enables us to detect the virtual states via solving eigenvalue problems, which are challenging in the conventional CSM. For the coupled-channel problems, one can choose different contours for each channel to detect the target poles in different regions of various Riemann sheets. 

To showcase its efficiency, we have applied this method to various potentials, such as the separable potential, the HAL QCD $DD^*$ potential, and the high-quality $NN$ Reid93 potential and compare our ICSM method with the LSE and CSM. We find that the ICSM can capture the poles that are undetectable by the conventional CSM. The ICSM yields the same results as the LSE through a more tractable process of solving an eigenvalue problem with a much faster speed and less computation time and cost. The ICSM can be employed to deal with  many elusive near-threshold signals in atomic physics, nuclear physics and, high-energy physics.

\clabel[ICSMvsCSMp2]{We also emphasize that this does not imply the CSM should be replaced. The simple transformation of the integral path, $p \rightarrow p e^{-i \theta}$, is evidently easier to implement. Therefore, when dealing with some complicated systems, the CSM remains an efficient and convenient method, especially when virtual states are not a primary concern. For example, when solving few-body systems, considering the significantly increased number of momenta to be integrated, dealing with a flexible integral path for each momentum would be challenging. On the other hand, the transformation $p_i \rightarrow p_i e^{-i \theta}$ is  straightforward.}

\section*{ACKNOWLEDGMENTS}
We are grateful to Jian-Bo Cheng, Yao Ma, Jun-Zhang Wang, Liang-Zhen When, and Wei-Lin Wu for the helpful discussions. This project was supported by the National
Natural Science Foundation of China (11975033 and 12070131001). This
project was also funded by the Deutsche Forschungsgemeinschaft (DFG,
German Research Foundation, Project ID No. 196253076-TRR 110). 

\bibliography{references}
\end{document}